\documentclass{PoS}

\usepackage{url}

\title{Multi-Messenger Observations of GRBs: \\ The GW connection}

\ShortTitle{GRBs and GWs}

\author{\speaker{Elisabetta Bissaldi} \\
        Dipartimento Interateneo di Fisica, Politecnico di Bari, Via G. Amendola 126, 70125 Bari, Italy\\
        Istituto Nazionale di Fisica Nucleare (INFN) - Sezione di Bari, Via E. Orabona 4, 70125 Bari, Italy\\
        E-mail: \email{elisabetta.bissaldi@ba.infn.it}}

\abstract{Two years ago, the astronomical community witnessed a historical breakthrough observation: the detection of a short Gamma-Ray Burst (GRB) by gamma-ray instruments in coincidence with the detection of a Gravitation Wave (GW) signal produced by the coalescence of two binary neutron stars. This joint GRB-GW observation paved the way to a new chapter in modern astrophysics: the "Multi-Messenger" era. In this contribution, I will review the main results by gamma-ray experiments obtained from 2015 to 2017 during the first two observational runs of the LIGO-Virgo experiments (O1/O2), and highlight strategies and status during the current run (O3) which will cover a 1-year period starting April 2019. Finally, I will focus on future prospects for gamma-ray missions dedicated to GW counterpart studies.}

\FullConference{36th International Cosmic Ray Conference -ICRC2019-\\
		July 24th - August 1st, 2019\\
		Madison, WI, U.S.A.}

\begin{document}

\section{Introduction}
Since their serendipitous discovery in the late 1960's, Gamma-Ray Bursts (GRBs) have represented an intriguing astrophysical phenomenon. GRBs are short and sudden electromagnetic (EM) signals in the gamma-ray band which, for a few blinding seconds, become the brightest objects in the known Universe. On average, they are observed more than once per day. For more than 20 years following the publication of their discovery \cite{1973ApJ...182L..85K}, GRBs were basically studied in the gamma-ray regime, namely between few keVs and few MeVs, where the bulk of their emission was detected by several dedicated space missions. However, GRBs remained only roughly localized and vanished too soon, leaving no traces: due to the large error boxes of the positions given by the gamma-ray instruments, no GRB counterpart could be detected at any other wavelength.

A first major step forward in understanding the GRB phenomenon was possible thanks to the instruments on-board the Compton Gamma-Ray Observatory (CGRO, 1991-2000), and in particular the Burst and Transient Source Experiment (BATSE), which detected 2704 bursts over a 9-year period \cite{1999ApJS..122..465P}. BATSE's sensitivity and good time resolution allowed the study of the detailed energy and temporal evolution of the GRB spectra, thus unveiling important information about the nature of GRBs. This included (i) their isotropical distribution across the sky, strongly suggesting a cosmological origin; (ii) the bimodal distribution of the burst durations, introducing two classes of events, the long (duration $>2$\,s) and the short (duration $<2$\,s) GRBs \cite{1993ApJ...413L.101K}; (iii) the irregularity and variability of their lightcurves, providing strong justification for the compact size of the burst emitting region; and (iv) the spectral properties, showing that GRBs basically emit non-thermal radiation. 
The first hint to the multi-wavelength nature of GRBs was given thanks to the detection by the EGRET and COMPTEL instruments onboard CGRO of a high-energy population of bursts, which emitted most of the energy in the MeV range. Some events were detected by EGRET at energies $>100$\,MeV \cite{1995Ap&SS.231..187D}, others displayed high-energy emission with large temporal delays with respect to the trigger time \cite{1994Natur.372..652H}. The production of high-energy photons and the relation between high and low-energy emission could not be fully understood at the time. 
During the CGRO era, a major goal that the GRB community wanted to achieve consisted in obtaining the GRB position while it was still bursting, in order to transmit that information to instruments capable of making rapid follow-up observations. Therefore, the BACODINE system was introduced  \cite{1995Ap&SS.231..235B}, which was capable of distributing BATSE locations to sites and instruments around
the world, greatly reducing the time delays from days to seconds. As soon as other missions/instruments were added, the name was changed from BACODINE to the more general {\it Gamma-ray Coordinates Network}\footnote{\url{https://gcn.gsfc.nasa.gov/}} (GCN) name still used at present. Another important help was the collaboration among operating gamma-ray detectors through the {\it Interplanetary Network}\footnote{\url{http://www.ssl.berkeley.edu/ipn3/index.html}} (IPN), aiming to infer precise burst locations by timing the arrival of the GRB signal at several spacecrafts \cite{2017ApJS..229...31H}. An important IPN member since the early times is the Konus-Wind experiment (launched 1994).

One of the most important observational breakthroughs in the study of GRBs was due to the discoveries by the Italian-Dutch satellite Beppo-SAX (1996-2003). Its innovative capabilities allowed to determine on-board a burst position with arc-minute precision and to slew its narrow-field X-ray telescopes towards that position. This first real-time follow-up uncovered the so-called ``afterglow'' emission \cite{1997Natur.387..783C}, which occurred at lower energies (from X-ray to optical to radio wavelengths) and on much longer timescales, lasting for days to months after the initial prompt gamma-ray emission. In some cases, ground-based observations revealed a close proximity of the afterglow to the optical light of faint galaxies, which were later identified as the ``host'' galaxies of the GRB progenitors \cite{1997Natur.386..686V}. These findings finally allowed precise measurements of GRB distances, firmly confirming them as cosmological sources. Moreover, a standard model explaining the GRB emission mechanisms began to take shape: in this picture, a highly relativistic jet is powered by matter interacting with a compact object such as an accreting black hole (BH) or a magnetar. This so-called central engine should be produced by the core-collapse of a rapidly rotating massive star or the merger of two compact objects, such as binary neutron stars (BNS) or a neutron star-black hole (NS-BH) system. The former scenario is traditionally attributed to long (duration $>2$\,s) GRBs, where supernova signatures have been seen in some late-time spectra \cite{2001ApJ...550..410M}. On the other hand, merging compact objects are the prime candidate to power short GRBs \cite{1989Natur.340..126E}. Other space missions which were launched shortly thereafter were the HETE-2 (2000-2008) and the INTEGRAL (launched  2002) satellite. While the first was dedicated to multi-wavelength observation of GRBs with X- and gamma-ray instruments, the latter is still exploring the gamma-ray sky for various high-energy sources. Particularly interesting for GRB-related studies is the anti-coincidence shield (ACS) of its spectrometer SPI. Although SPI-ACS does not have an imaging capability, it serves as a large effective area detector with a quasi-omnidirectional field of view \cite{2012A&A...541A.122S}. 

The GRB afterglow discovery officially opened the way to the  multi-wavelength study of the GRB phenomenon. In this context, a major player since its launch in 2004 has been the {\it Neil Gehrels Swift Observatory}, which thanks to its three instruments, BAT, XRT and UVOT, covers a broad energy range, thus providing arc-second positions for precise follow-up on very short timescales. 
Another major breakthrough came in 2005, with the identification of the first host galaxies of short GRBs and multi-wavelength observation of their afterglows \cite{2007PhR...442..166N}.
%(Berger et al. 2005; Fox et al. 2005; Gehrels et al. 2005; Hjorth et al. 2005b; Villasenor et al. 2005). %for reviews see Nakar 2007; Berger 2014)  
These observations strongly hinted to short GRBs being associated with BNS or NS-BH mergers, yet proofs remained elusive. However, merging compact objects were seen as a potential source of gravitational waves (GWs) and interest intensified in following up GW detections electromagnetically \cite{2012ApJ...746...48M}.
Other important missions following Swift were Suzaku (2005-2015), AGILE (launched 2007) and Fermi (launched 2008), which although not primarily dedicated to GRB studies, greatly enhanced the GRB knowledge. In particular, the Fermi spacecraft carries two instruments onboard, the Large Area Telescope (LAT), covering energies from 100 MeV to $>300$\,GeV, and the Gamma-Ray Burst Monitor (GBM), which, like its predecessor BATSE, is specifically designed for detecting GRBs over its full-sky field of view in an energy range between 8 keV and 40 MeV, and has collected more than 2600 GRBs so far \cite{2016ApJS..223...28N}.
The latest gamma-ray instruments collecting valuable GRB data include MAXI-GSC (since 2009), ASTROSAT-CZTI (since 2015), Insight-HXMT (since 2017), and CALET-GRBM (since 2015).
The multi-wavelength studies of GRBs have recently expanded also into the very high energy (VHE) domain, with H.E.S.S announcing late-time VHE emission from GRB\,180720B and MAGIC detecting VHE emission from GRB\,190114C. These discoveries have truly shown that GRBs need to be studied across the entire EM spectrum. Given the large amounts of energy released in these events, they are also a natural candidate for studies in other domains, such as neutrino emission and GWs.
\begin{figure}[t!]
\centering
\includegraphics[width=0.7\columnwidth]{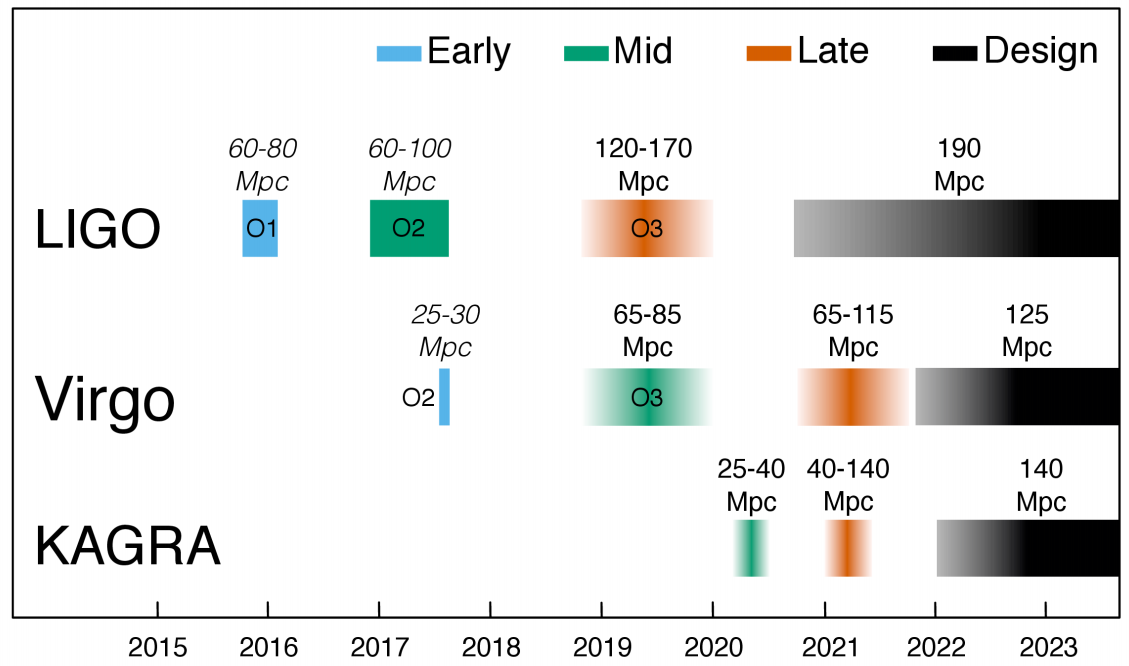}
\caption{The planned sensitivity evolution and observing runs of the aLIGO, AdV and KAGRA detectors over the coming years. There is significant uncertainty in the start and end times of future observing runs, and these could move forward or backwards relative to what is shown above. From \cite{2018LRR....21....3A}.}
\label{fig_4}
\end{figure}
\section{The GW connection}
Several theories proposed since almost 30 years illustrate how in the compact binary coalescence model of short GRBs, a NS and a compact companion in otherwise stable orbit lose energy to gravitational waves and spiral in. The neutron star(s) tidally disrupt shortly before coalescence, providing matter, some of which is ejected in relativistic jets, thus powering the prompt emission seen by gamma-ray instruments \cite{2009ARA&A..47..567G}.
% Eichler et al. 1989; Narayan et al. 1992; Nakar 2007; Gehrels et al. 2009
This tight connection bewteen short GRBs and GW-emitting sources brought the GRB and GW communities closer together, having a possible joint detection of these objects as common goal. Moreover, an EM counterpart discovered through a follow-up of a gravitational wave candidate event was considered as a fundamental increase in the confidence of the astrophysical origin of the GW signal, while also providing complementary insights into the progenitor and environment physics.

The research and development of GW detectors started back in the 70's, and led to the construction of the major currently-operating ground-based laser interferometers in the 90's. These comprise (1) the pioneering GEO600, located near Hanover (Germany); (2) the Laser Interferometer Gravitational-Wave Observatory (LIGO), consisting of two widely separated identical detector sites in the US working as a single observatory (Hanford in Washington State and Livingston in Louisiana); and (3) Virgo, an interferometer located outside of Pisa (Italy), funded by the European Gravitational Observatory (EGO). While all these collaborations are separate organizations, they cooperate closely and are referred to as LVC. A fourth interferometer, the KAmioka GRavitational-wave Antenna (KAGRA), located in Japan, will join the GW network soon: it is currently being upgraded to its baseline design configuration and will be operational in its full configuration by the end of 2019.

Initial LIGO took data between 2001 and 2010, almost contemporary with initial Virgo, without detecting any GW signal. Several searches for  coalescence signals or unmodeled GW bursts  associated with GRBs were performed during so-called "science runs" of both detectors \cite{2010ApJ...715.1438A}.
% Abbott et al. 2005, 2008b; Acernese et al. 2007, 2008,  % Abbott et al. 2010b, 2010a.
Moreover, the development of low-latency GW data analysis pipelines allowed the use of GW candidate signals to conduct EM follow-up programs \cite{2012ApJ...760...12A}. Early follow-up programs involved ground-based and space EM facilities observing the sky in different EM bands, mainly in the optical and radio bands, with the Swift satellite being the only space-based telescope involved at the time.

The GW follow-up program represented a milestone toward the advanced detector era: LIGO and Virgo went through several years of redesign, construction, preparation and installation, which finally led to Advanced LIGO (aLIGO) and Advanced Virgo (AdV) being ready to start operations in 2015 and 2017, respectively. The improvements made both observatories 10 times more sensitive, allowing to increase the volume of the observable universe by a factor of 1000. 
\begin{figure}[t!]
\centering
\includegraphics[width=0.95\columnwidth]{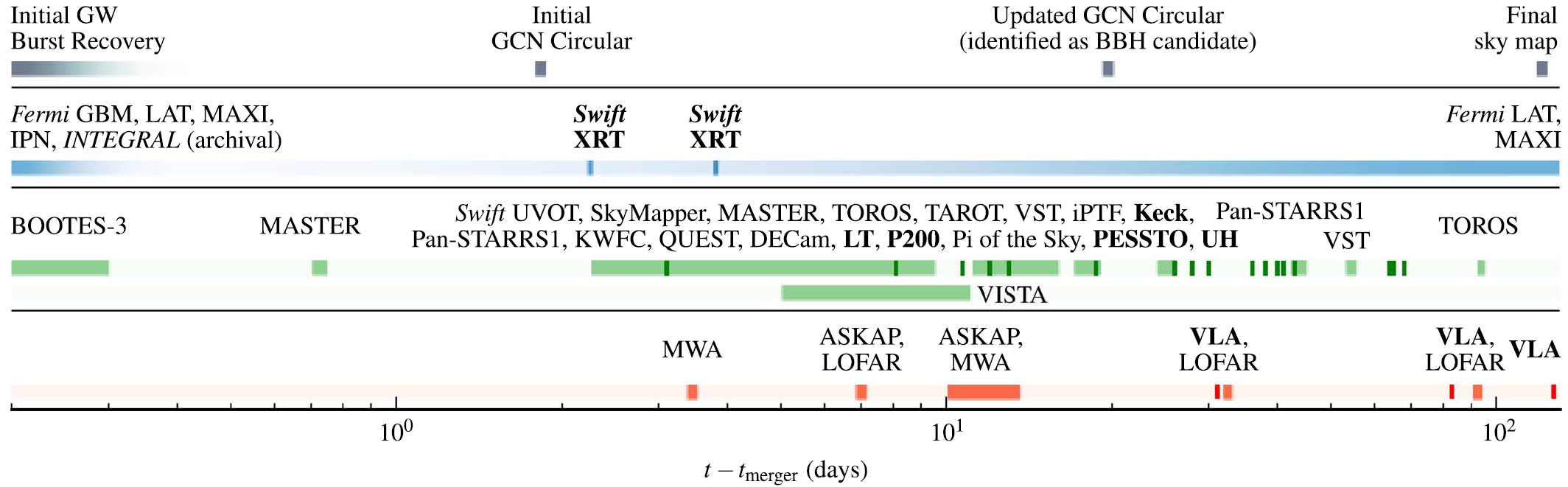}
\caption{Timeline of observations of GW150914, separated by band and relative to the time of the GW trigger. The top row shows GW information releases. The bottom four rows show high-energy, optical, near-infrared, and radio observations, respectively. From \cite{2016ApJ...826L..13A}.}
\label{fig_6}
\end{figure}
\section{Start of the GW/EM multimessenger era}
In September 2015, aLIGO began the era of GW astronomy with its first observation run (O1) and detections, collecting data until January 2016. During this run (as well as during the second one, O2, November 2016--August 2017), exchange of GW candidates with partners outside LVC was governed by memoranda of understanding (MoU). %; Abadie et al 2012d; Aasi et al 2013b). 
The interferometers were not yet operating at design sensitivity during O1, achieving a 60--80 Mpc BNS range (see Fig.\,\ref{fig_4}). 
Data from GW detectors were searched for many types of possible signals, in particular from compact binary coalescences (CBCs) and generic transient or burst signals. CBCs include BNS, NS--BH and BBH systems.
During O1, aLIGO reported two clear detections, GW150914 \cite{2016PhRvL.116f1102A} and GW151226 \cite{2016PhRvL.116x1103A}, and a lower significance candidate, LVT151012 \cite{2016PhRvX...6d1015A}, recently confirmed as a confident detection (and thus renamed GW151012). All three events originated from BBH coalescences. As part of the EM follow-up program, the times and sky localizations were promptly distributed using the GCN system. For GW150914, 63 teams participating in the MoU were operational and covered the full EM spectrum. Thorough checks were performed before the GW alert was released, resulting in latencies much greater than 1 h and in 25 teams responding to it, spanning 19 orders of magnitude in EM wavelength. The chronology of the GW detection alerts and follow-up observations of GW150914 is shown in Fig.\,\ref{fig_6}.  No significant EM counterpart and no afterglow emission was found in optical, UV, X- or gamma-rays \cite{2016ApJ...826L..13A}.  Also no significant neutrino emission was found to be temporally and spatially coincident with the event. Nevertheless, this first broadband campaign represented a milestone, highlighting the broad capabilities of the transient astronomy community and the observing strategies that have been developed to pursue BNS merger events. In fact, no EM emission is expected for vacuum BBH mergers \cite{2010RvMP...82.3069C}, but it is possible if there is surrounding material, for example, remnants of mass lost from the parent star \cite{2016ApJ...821L..18P} or if the binary was embedded in a circumbinary disc or a common envelope \cite{2016ApJ...824L..10W}.
\begin{figure}[t!]
\centering
\includegraphics[width=0.98\columnwidth]{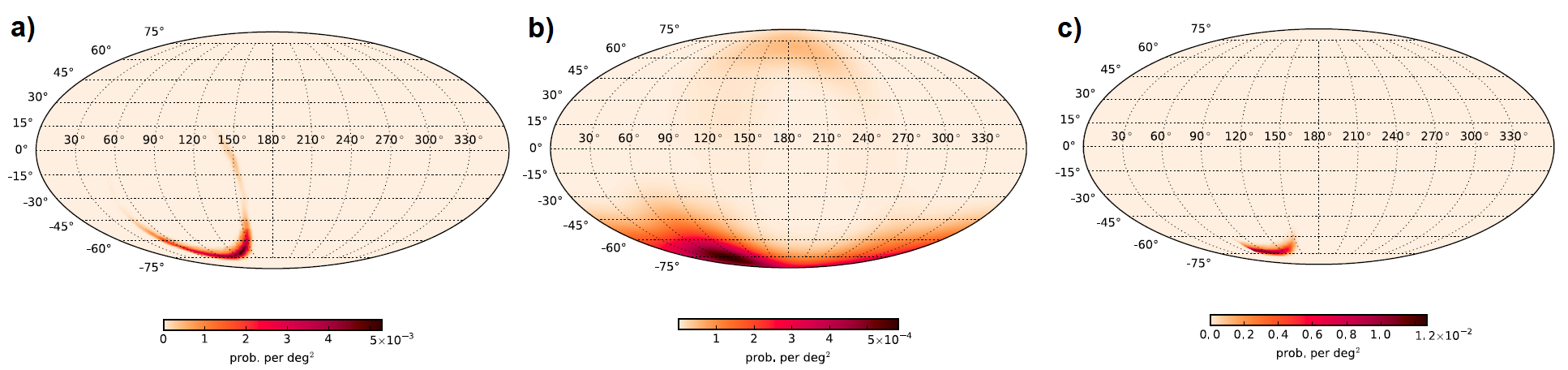}
\caption{Localization maps of GW150914: (a) LIGO; (b) GBM; (c) combined LIGO/GBM, in case of confirmed association. Figure adapted from \cite{2016ApJ...826L...6C}.}
\label{fig_7}
\end{figure}
\subsection{The curious case of GW150914-GBM}
Fermi-GBM reported the detection of a weak but hard transient occurring 0.4\,s after the LIGO GW trigger GW150914, with a false alarm probabilty corresponding to $2.9\,\sigma$ \cite{2016ApJ...826L...6C}. This signal, referred to as GW150914-GBM, was detected through dedicated offline searches (so-called targeted searches) of the GBM data, developed by GBM/LIGO team members for retrieving events too weak to trigger GBM on board, non-triggered short GRBs in particular. The GBM team performed extensive analysis of the detected transient to address its astrophysical origin. The source direction was calculated to lie underneath the Fermi spacecraft, i.e. was highly unfavourable for the GBM detectors and resulted in a very large uncertainty region (see Fig.\,\ref{fig_7}, panel b), yet still consistent with the southern lobe of the LIGO localization annulus (as shown in Fig.\,\ref{fig_7}, panel a). The spectrum of GW150914-GBM extended into the MeV range, revealing fit parameter values and a fluence ($2.4 \times 10^{\,-7}$\,erg cm$^{\,?2}$) which are average for short GRBs. Moreover, the nature of GW150914-GBM was found incompatible with galactic transient sources, solar and terrestrial emission. However, a magnetospheric origin could not be totally excluded, even if the observing conditions were not conducive to such an event, nor was the light curve typical of such activity.

GW150914-GBM raised a lot of discussion among the astrophysical community. In particular, the non-detection by other space-based instruments like AGILE and INTEGRAL SPI-ACS was regarded as indicative of the transient not being associated to the GW signal. These non-detections were further discussed in \cite{2018ApJ...853L...9C}, resulting in a statement that any tension between the GBM and other instruments would have been resolved only with future joint observations of GW events and by closer collaborations of the various teams. In any case, Fermi-GBM still proves to be an ideal partner in the search for EM  signals in coincidence with GW detections, thanks to its broad field of view and good sensitivity at the peak emission energies for short GRBs. Moreover, as shown in panel c) of Fig.\,\ref{fig_7}, even a large uncertainty region provided by GBM combined with the LIGO annuls (in case of confirmed association) would help shrinking the LIGO localization by almost $2/3$.
Other dedicated Fermi-GBM and Fermi-LAT searches for EM counterparts were performed to GW151226 and LVT151012 \cite{2017ApJ...835...82R}, yet no EM counterpart was found detected. In fact, these non-detections can neither confirm nor refute the potential association between GW150914 and the GBM candidate counterpart. However, putting these GW detections in the context of the more familiar short GRBs already demonstrated the capability of both the GBM and LAT for future searches.
Recently, \cite{2019ApJ...871...90B} presented a comprehensive search for prompt gamma-ray counterparts to all CBC candidates collected during O1. These CBC candidates were well below the standards for the GW trigger alone to be considered likely due to a compact merger, but significant enough that if a gamma-ray transient was found by GBM in coincidence, it would support an astrophysical origin of the GW transient. GW candidate event times were followed up by targeted searches of Fermi-GBM data. Among the resulting GBM events, the two with the lowest false alarm rates were the GW150914-GBM and a solar flare in chance coincidence with a GW candidate. Conversely, no GW candidates were found to be coincident with gamma-ray transients independently identified by blind searches of the GBM data. 
Many more searches for EM signals possibly coincident with O1 GW triggers have been conducted by other gamma-ray instruments, including INTEGRAL \cite{2016ApJ...820L..36S}, 
%Savchenko et al. ApJL 820 L36 2016
%Mereghetti et al. 2018Memorie della Societa Astronomica Italiana, v.89, p.230 (2018) 
AGILE \cite{2016ApJ...825L...4T}, and
%Tavani et al., ApJL 825 L4 2016
%Verrecchia et al. 2018 Proceedings of the International Astronomical Union, IAU Symposium, Volume 338, pp. 84-89
Swift \cite{2016MNRAS.462.1591E},
%Evans et al. 2016, MNRAS 460, L40-L44
%EVans et al. 2016, MNRAS 462, 1591-1602
leading to no detections.
\begin{figure}[t!]
\centering
\includegraphics[width=0.7\columnwidth]{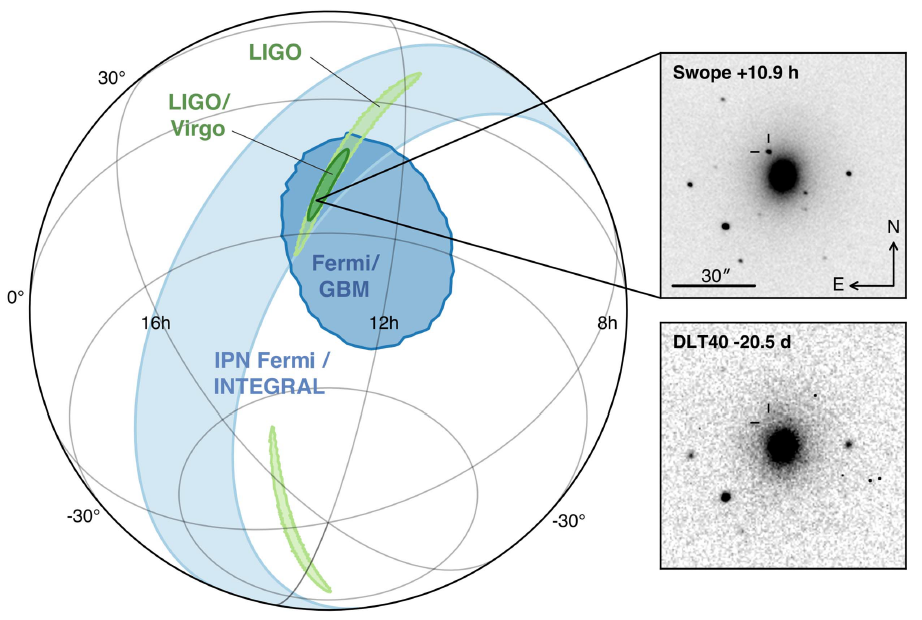}
\caption{Localization of the gravitational-wave, gamma-ray, and optical signals for GW170817/ GRB\,170817A. The inset shows the location of the apparent host galaxy NGC 4993 in the Swope optical discovery image (top right) and the DLT40 pre-discovery image. The reticle marks the position of the transient in both images. From \cite{2017ApJ...848L..12A}.}
\label{fig_10}
\end{figure}
\subsection{GW170817 / GRB 170817A}
The second observing run (O2) of aLIGO started at the end of November 2016 (see Fig.\,\ref{fig_4}). AdV joined the O2 run at the beginning of August, 2017. Both teams ended O2 operations on August 25, 2017. During this time, LVC detected 8 confirmed GW events. Together with the three GW events from O1, these events make up the first {\it GW Transient Catalog of Compact Binary Mergers Observed by LIGO and Virgo during the First and Second Observing Runs} (GWTC-1) \cite{2018arXiv181112907T}. 
A first event worth mentioning is the BBH event GW170814, which represented the first joint LIGO-Virgo detection \cite{2017PhRvL.119n1101A}. It proved that a three-detectors network can improve the sky localization of the source and reduce the GW annulus region by $>90\%$, thus greatly facilitating EM follow-up programs. 

The first direct evidence of a link between BNS mergers and short GRBs came only three days later, on August 17, 2017 \cite{2017PhRvL.119p1101A}. On that day, Fermi-GBM triggered on, classified, and localized a short burst, GRB 170817A,  automatically announcing it trough GCN $\sim$15\,s later. At the same time, a GW candidate (later designated GW170817) was registered in low latency in LIGO-Hanford data. The signal was consistent with a BNS coalescence with merger time less than 2\,s before GRB 170817A. A GCN Notice was issued $\sim$25\,min later. In the first hours after the initial GW detection, GW data from LIGO and Virgo were combined to produce a three-instrument skymap, placing the source nearby at a luminosity distance of $\sim$40\,Mpc, in an elongated region of $\sim$28 deg$^2$ on the sky (dark green contour in Fig.\,\ref{fig_10}). The GBM location resulted to be broadly consistent with one of the quadrupole lobes from the skymap produced by the GW interferometers (dark blue contour in Fig.\,\ref{fig_10}). Moreover, a follow-up search for short transients in SPI-ACS data identified a single excess, which was then associated to the GBM observation of GRB\,170817A with a significance of $4.2\sigma$ \cite{2017ApJ...848L..13A}. Exploiting the difference in the arrival time of the gamma-ray signals at GBM and SPI-ACS provided additional significant constraints on the gamma-ray localization area (light blue stripe in Fig.\,\ref{fig_10}). Other gamma-ray missions such as Insight-HXMT, CALET, AGILE and Fermi-LAT searched their data for possible associated signal, without detecting any significant excess  \cite{2017ApJ...848L..12A}. At even higher energies, H.E.S.S. and HAWC also performed follow-up campaigns, revealing no significant gamma-ray emission.

The announcements by GBM and LVC, together with the well-constrained localization, triggered one of the largest broadband observing campaigns thus far, using both ground- and space-based telescopes. The Swope telescope at Las Campanas Observatory in Chile detected a bright optical transient $\sim10$\,hrs after the GW trigger, later designated SSS17a/AT 2017gfo, located at a distance consistent with the GW-inferred one (see inset in Fig.\,\ref{fig_10}). %Coulter et al. (2017) 
Over the subsequent weeks, a network of observatories spanning the UV, optical and near-IR wavelengths followed up GW170817. The evolution of the spectral energy distribution, rapid fading, and emergence of broad spectral features led to the conclusion that the source had physical properties matching kilonovae theoretical predictions \cite{2017ApJ...848L..19C}.
% (Chornock et al. 2017c; Kasliwal et al. 2017, Covino et al. 2017)

\begin{figure}[t!]
\centering
\includegraphics[width=0.8\columnwidth]{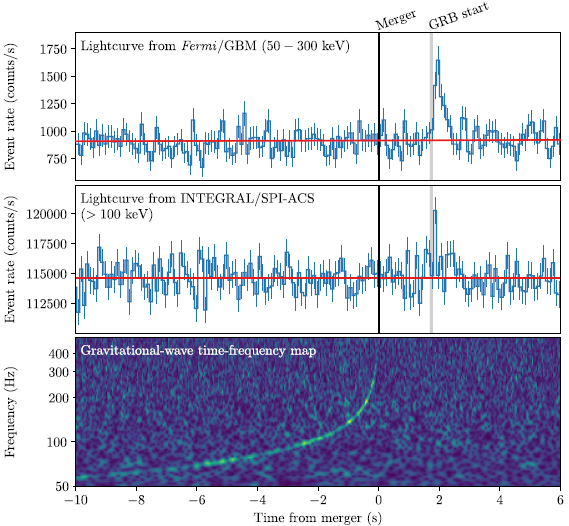}
\caption{Multi-messenger detection of GW170817 and GRB 170817A. From top to bottom: Summed GBM lightcurve ($50-300$ keV); SPI-ACS lightcurve ($>100$ keV); Time-frequency map of GW170817. All times are referenced to the GW trigger time. Adapted from \cite{2017ApJ...848L..13A}.}
\label{fig_11}
\end{figure}

\begin{figure}[t!]
\centering
%\vspace{10pt}
\begin{tabular}{cc}
\includegraphics[height=0.31\columnwidth]{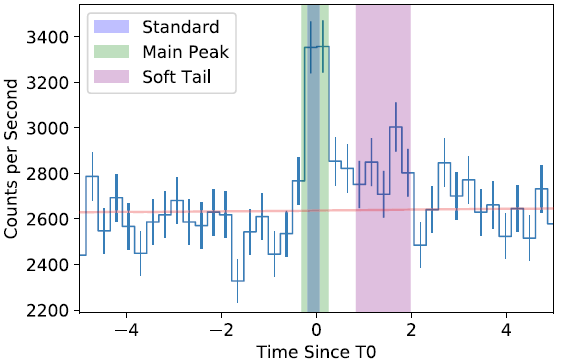} &
\includegraphics[height=0.32\columnwidth]{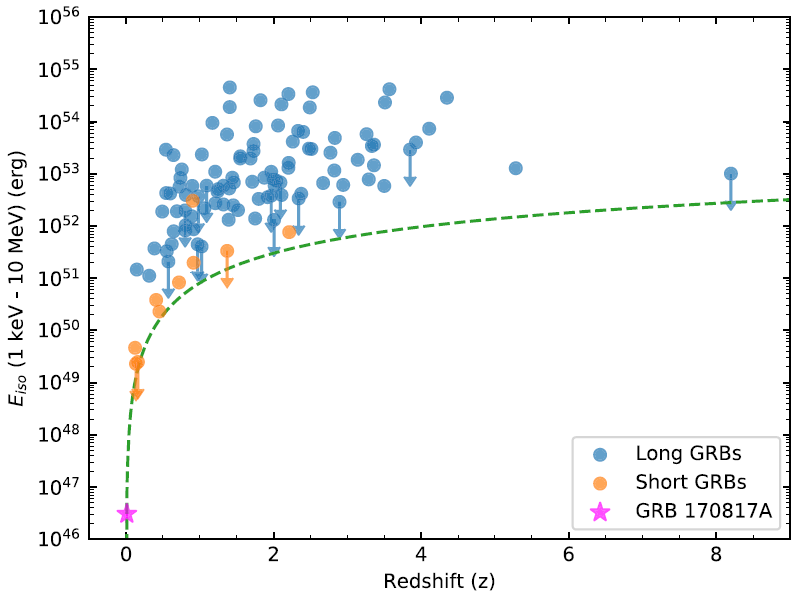}
\end{tabular}
\caption{Left panel: Fermi-GBM light curve of GRB\,170817A in the $10-300$ keV band. Shaded regions mark different time intervals selected for spectral analysis.  From \cite{2017ApJ...848L..14G}. Right panel: Distributions of the isotropic emitted energy as a function of redshift for all GBM-detected GRBs with measured redshifts. The green curve demonstrates how the (approximate) GBM detection threshold varies as a function of redshift. From \cite{2017ApJ...848L..12A}.}
\label{fig_8}
\end{figure}

Fig.\,\ref{fig_11} shows the first ever multi-messenger ``lightcurve'' of a BNS merger, from 10\,s before to 6\,s after the GW trigger. The time-frequency representation of the LIGO data containing GW170817 is shown in the bottom panel, while the top panels show the summed GBM lightcurve in the $50-300$ keV energy range and the SPI-ACS lightcurve at energies $>$100 keV (with a high-energy limit of least 80 MeV). The inferred time delay between the GW and the gamma-ray signal was of $1.74 \pm 0.05$\,s. These joint observations made it possible to set stringent limits on fundamental physics, i.e., proving that gravity and light travel at the same speed, and to probe the central engine of short GRBs in ways that have not been possible with EM data alone.
Despite GRB 170817A being the closest short GRB with measured redshift ($z \sim 0.008$) so far, it was very subluminous. Detailed analysis of Fermi-GBM data revealed two emission components: a main pulse lasting $\sim0.50$\,s, followed by a weak tail extending to almost 2\,s post trigger (see Fig.\,\ref{fig_8}, left panel). The two components showed different spectral properties, with the latter exhibiting a softer, blackbody spectrum \cite{2017ApJ...848L..14G}. 
Using spectral information and the distance of the burst, the inferred energetics of GRB 170817A clearly indicated that it is 2 to 6 orders of magnitude less energetic than other short GRBs. This is particularly evident when comparing this burst with other GBM-detected bursts with redshift measurements (see Fig.\,\ref{fig_8}, right panel). Several explanations were proposed for the observed dimness of the burst, mainly involving a mixture of an intrinsic brightness distribution and geometric effects. Recently, very late-time (after $>200$\,days post trigger) radio observations 
showed that the burst dimness might have been a consequence of the burst jet type, which can be described as ``structured'' \cite{2019Sci...363..968G}. %Ghirlanda et al. Science 363, 968 2019.

The historical observation of GW170817/GRB\,170817A has proven that increased detection confidence, improved sky localization, and identification of host galaxy and redshift are fundamental benefits of joint GW-EM observations. A global network of GW detectors and wide-field gamma-ray instruments, such as Fermi-GBM and INTEGRAL/SPIACS, are critical to the future of multi-messenger astronomy in the GW era. Finally, one important aspect discussed in GWTC-1 derived from the complete analysis of all O1+O2 results is the calculation of the astrophysical merger rate densities of BBH, BNS, and NSBH systems in the local Universe. These supersede earlier estimates and limits from previous LVC results, since they involve more sophisticated data treatment in light of the confident detection of 1 BNS and 10 BBH mergers. The inferred rate density of BNS lies between 110 and 3840\,Gpc$^{\,-3}$\,yr$^{\,-1}$.
% - BBH mergers: 9.7--101 \,Gpc$^{\,-3}$\,yr$^{\,-1}$
% - NSBH binaries: upper limit 610 \,Gpc$^{\,-3}$\,yr$^{\,-1}$
The rates per unit volume of NSBH and BBH mergers are lower than for BNSs, but the distance to which they can be observed is larger. Consequently, the predicted observable rates are comparable.
%(Abadie et al. 2010b; Rodriguez et al. 2015; Abbott et al. 2016b; Li et al. 2017).
During O2 (9-months run) the expected aLIGO range was 80-120 Mpc, and the achieved range was in the region of 60-100 Mpc, while the expected AdV range was 20-65 Mpc.
\section{Current status and future prospects}
As of writing, GRB~170817A remains the only GRB reliably connected with a GW signal. However, observations are ongoing and each GW alert is checked for possible EM counerparts. The third observing run (O3) of aLIGO and AdV started on April 1st, 2019, and it is planned to last for 12 months (see Fig.\,\ref{fig_4}). Improvements achieved in the sensitivity of the detectors and the fact that the three LIGO-Virgo instruments have been operating simultaneously since day one, are now enabling unprecedented opportunities. In addition, for the first time, LIGO and Virgo are providing public alerts. These are delivered shortly after the detection of credible transient GW candidates. This strategy aims to facilitate follow-up observations by other telescopes and enhance the extraordinary potential of multi-messenger observations. Moreover, public web pages reporting the live status of the LIGO/Virgo detectors and alert infrastructure comprise the {\it Detector Status Portal}\footnote{\url{https://www.gw-openscience.org/detector\_status/}}, summarizing daily detector performances, the {\it GWIStat}\footnote{\url{https://ldas-jobs.ligo.caltech.edu/~gwistat/gwistat/gwistat.html}} or the {\it LIGO Data Grid Status}\footnote{\url{https://monitor.ligo.org/gwstatus}} pages, with real-time detector up/down status.

As of July 16th, 2019, 21 LVC public alerts have been issued since the beginning of O3. All alerts are freely accessible through the {\it Gravitational Wave Candidate Event Database}\footnote{\url{https://gracedb.ligo.org/}}. All but one events were classified as candidate BBH mergers. Full assessment requires more analysis, which is already ongoing. If confirmed, these candidates would add to the catalogue of 10 BBH mergers detected in previous runs.

%Currently, GW searches can also be triggered by EM observations, particularly by GRBs. Other possible targets for these externally-triggered GW searches include EM or neutrino emission from Galactic core-collapse supernovae. Therefore it is important to have telescopes capable of observing the high-energy spectrum operating during the advanced-detector era (and beyond). Furthermore, the rapid identification of a GW counterpart to such a trigger could prompt further spectroscopic studies and longer, deeper follow-up in different wavelengths that may not always be done in response to GRBs. This is particularly important for bursts with larger sky localization uncertainties, such as those reported by the Fermi-GBM, which are not followed up as frequently as bursts reported by Swift or Fermi-LAT.

Three events collected during O3, S190425z\footnote{\url{https://gracedb.ligo.org/superevents/S190425z/view/}}, S190426c\footnote{\url{https://gracedb.ligo.org/superevents/S190426c/view/}} and S190510g\footnote{\url{https://gracedb.ligo.org/superevents/S190510g/view/}}, point to coalescences of binary systems involving at least one NS. S190425z was four times more distant than GW170817 and the sky localization provided by LVC was much more uncertain, since the GW trigger occurred while only two detectors were operating. The second candidate signal, S190426c, was even fainter, with only a 49\%\ probability of being a BNS, with a remaining probability that it may not be a genuine astrophysical signal. Similarly, S190510g was reclassified to have a 58\%\ probability of being terrestrial and a 42\%\ of being a BNS. No EM counterpart was reported for any of these candidate events by other space- or ground-based observatories. Deeper investigations will be needed to uncover their true nature.

\begin{table}[t!]
\centering
\includegraphics[width=0.8\columnwidth]{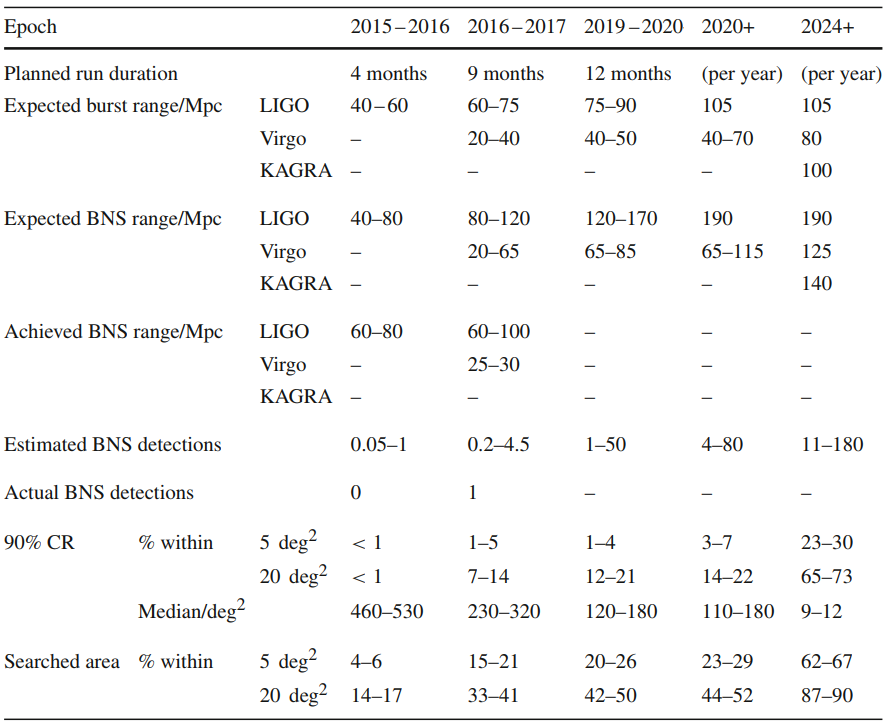}
\caption{Summary of a plausible observing schedule, expected sensitivities, and source localization with aLIGO, AdV and KAGRA detectors, which will be strongly dependent on the detectors' commissioning progress. From \cite{2018LRR....21....3A}.}
\label{tab_1}
\end{table}

In view of further sensitivity upgrades to both LIGO and Virgo as well as the prospects of KAGRA joining the network towards the end of 2019, many tens of binary observations are anticipated in the coming years. Sensitivity projections for observing and localizing GW transients with aLIGO, AdV, and KAGRA are given in Tab.\,\ref{tab_1}. For BNS events, the median sky localization accuracy in terms of the 90\%\ credible area will be 120-180 deg$^2$. 12-21\%\ of BNS mergers will be localized to less than 20 deg$^2$.
%The BNS range and burst range (luminosity distance of detectable sources, averaged over sky position and source orientation) given the detectors' anticipated sensitivities are listed in the table below. We expect 1–10 BNS events over the course of O3. 
Prospects for future runs are also given. From 2020 onward, a four-detector network will benefit from aLIGO at full sensitivity and AdV at 65-115 Mpc, later increasing to design sensitivity of 125 Mpc. Currently, 2024 seems to be the earliest time LIGO-India could be operational. With a four- or five-site detector network at design sensitivity, one may expect improved sky localizations as well as a much larger fraction of coincident observational time windows.  Multi-messenger follow-up of GW candidates will help confirm GW candidates that would not be confidently identified from GW observations alone. The big challenge over the next years for these follow-up campaigns will be the large position uncertainties, with areas of many tens to thousands of square degrees. 
\subsection{Future gamma-ray instruments}
Detecting a GRB counterpart also puts demands on the EM observatories available for follow-up. As GRB~170817A showed, a potential counterpart may be weak, and the factors governing this are not yet fully understood. This is likely to become even more important as the distances to which merging neutron star systems can be probed by GW detectors increases. Because of the large uncertainties in localization and expected short duration of the EM signals, instruments with a wide field of view and/or rapid repointing capabilities will be at the core of the efforts for the foreseeable future. Among the active missions, the extension of the Fermi and Swift is of primary importance. Fermi-GBM currently detects more prompt short GRBs than all other active mission combined, while Fermi-LAT is one of few instruments capable of covering the MeV-GeV range. Conversely, Swift is unparalleled regarding fast response X-ray and UV coverage, allowing the precise localization required for ground-based follow-up.

%\nob{I don't know if this text should go here, but I thought some general introduction was needed. Also, I moved the parts about Swift and Fermi here instead, before talking about new missions. It's commented out below.}

The most critical wavelengths for detecting GRB prompt emission are keV-MeV $\gamma$-rays and X-rays, which require space-based observatories. Localization with gamma-ray instruments can occur in two ways: autonomous real-time prompt GRB localization by a single detector (which can be improved with follow-up by other instruments on the same spacecraft) or the detector's use in the IPN for timing annulus localizations. Instruments in Low Earth Orbit (LEO) require distant instruments for these annuli to be constraining, given the limited timing accuracy from short GRB observations.
Sources that require long-term EM monitoring cannot be studied by scintillation detectors, mainly because of poor targeting capabilities, but rather by coded masks, and are best studied with Compton and pair-conversion survey telescopes.
%Follow-up ground observations with current or expected missions (e.g. HAWC, CTA, VLA, SKA, ZTF, LSST) reliably cover TeV, optical, near-IR, and radio. 

\begin{table}[t!]
\centering
\includegraphics[width=0.95\columnwidth]{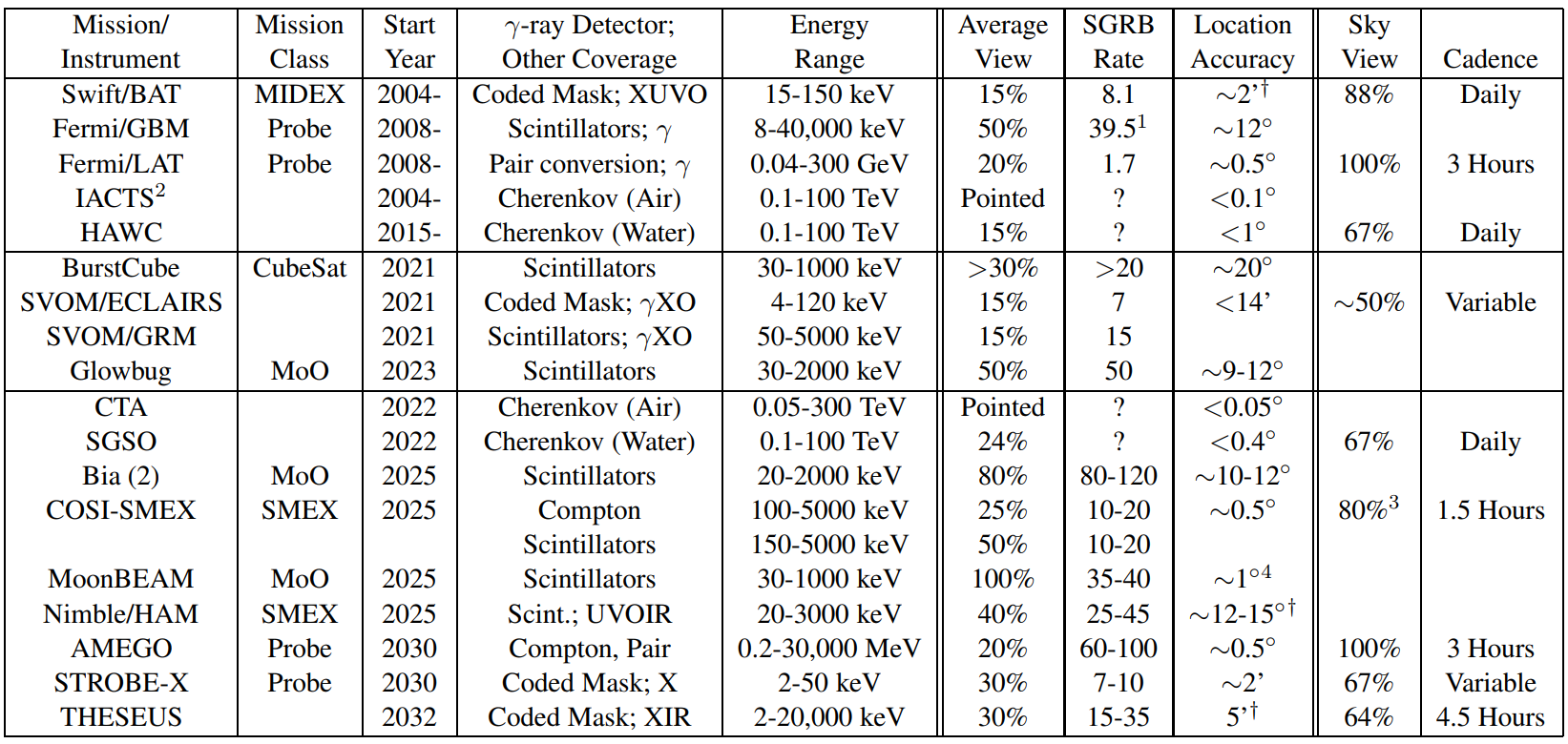}
\caption{A summary of representative active, funded and proposed gamma-ray missions. $\dagger$ denotes missions with on-board follow-up instruments which enable more accurate localizations. $^{1}$\,GBM also has subthreshold searches, identifying an additional $\sim$80 short GRB candidates/year; $^{2}$\,H.E.S.S./MAGIC/VERITAS; $^3$\,COSI-SMEX observes 100\%\ of the sky with a daily cadence; $^4$\,This localization assumes an IPN annulus in partnership with a LEO mission. Adapted from \cite{2019BAAS...51c.260B}.} %White paper Burns arxiv 1903.04472.pdf
\label{tab_2}
\end{table}

A summary of representative gamma-ray observatories is given in Tab.\,\ref{tab_2}, divided in three blocks: active (first block), funded (second block) and proposed (third block). These instruments are broadly classed as scintillators, coded masks, Compton, and pair-conversion telescopes and may be flanked by instruments observing other wavelengths (X, UV, O, and IR), as listed in column 4. 
Ground-based observatories are also listed. Higher energy photons are observed indirectly through Cherenkov radiation as the photons pass through water (in enclosed tanks, like for HAWC) or through the Earth's Atmosphere (with Imaging Atmospheric Cherenkov Telescopes, IACTs). All gamma-ray detectors are wide-field survey telescopes, except for IACTs. In Tab.\,\ref{tab_2}, the ``Average View'' column refers to the field of view modulo livetime, corresponding to the likelihood a given observatory will observe an event. The location accuracy is given in terms of 1\,$\sigma$ uncertainties (statistical and systematic). ``Sky View'' and ``Cadence'' refer to how often that fraction of the sky is observed. Estimated short GRB rates for Cherenkov telescopes are not given, since they are unknown. For future missions these estimates are based on the current design of the mission parameters and instruments and are subject to change. Representative values are assumed when not precisely known. References for all proposed instruments are given in \cite{2019BAAS...51c.260B}.

The ideal future detector for GW-GRB counterpart searches would be a large-scale gamma-ray observatory capable of detecting more short GRBs than prior missions, mainly through improved keV-MeV sensitivity and broad sky coverage. A capability to localize bursts to sufficient accuracy (ideally to arcsecond precision) would also be necessary for precise follow-up observations. These observations should then be matched with other follow-up facilities such as sensitive, fast-response, high spatial resolution X-ray telescopes and UV-O-IR and radio telescopes.
Apart from the instruments themselves, time-domain astronomy relies on rapid communication of alerts and results from data analysis. The challenges can be exemplified with the Fermi mission: although the GBM can provide an alert and approximate localization to the community within minutes, it takes hours before the LAT data reaches the ground and can be analysed for a potential counterpart. Developing the coordinated follow-up efforts thus requires improvements regarding on-board analysis as well as real-time communication for space-based missions. Possible technical improvements include real-time reporting, automated multi-mission and multi-messenger searches, and prompt reporting of initial parameter estimation from GW detections (e.g. masses) to enable follow-up prioritization. Extending the network to include neutrino observatories has already begun, paving the way to a true multi-messenger view of GRBs.

%Currently funded observatories include SVOM \nor{(REF)}, BurstCube \nor{(REF)} and Glowbug \nor{(REF)}.
%Future detectors which have been proposed comprise i.e. COSI, AMEGO and THESEUS. 
%\nob{I don't think you need to go into the future missions in detail so I commented it out. It's enough that you list a few examples in the table. They are anyway so far into the future that things will likely change! Also, I thought that the last part of this paragraph now makes it sound like a nice conclusion ;)}
%
%
%
%
\bibliographystyle{JHEP}
\bibliography{my_bib}
\end{document}